\documentclass[preprint,aps,superscriptaddress]{revtex4}

\usepackage{graphicx}
\usepackage{subfigure}
\usepackage{epsfig}
\usepackage{xcolor}
\usepackage{dcolumn}
\usepackage{bm}
\usepackage{ulem}
\usepackage{color}
\usepackage{amsmath}
\usepackage{amsfonts}
\usepackage{amssymb}
\usepackage{epstopdf}

\begin{document}

\title{Experimental observation of topological Fermi arcs in type-II Weyl semimetal MoTe$_2$}

\author{Ke Deng}
\altaffiliation{These authors contribute equally to this work.}
\affiliation{State Key Laboratory of Low Dimensional Quantum Physics and Department of Physics, Tsinghua University, Beijing 100084, China}

\author{Guoliang Wan}
\altaffiliation{These authors contribute equally to this work.}
\affiliation{State Key Laboratory of Low Dimensional Quantum Physics and Department of Physics, Tsinghua University, Beijing 100084, China}

\author{Peng Deng}
\altaffiliation{These authors contribute equally to this work.}
\affiliation{State Key Laboratory of Low Dimensional Quantum Physics and Department of Physics, Tsinghua University, Beijing 100084, China}

\author{Kenan Zhang}
\affiliation{State Key Laboratory of Low Dimensional Quantum Physics and Department of Physics, Tsinghua University, Beijing 100084, China}

\author{Shijie Ding}
\affiliation{State Key Laboratory of Low Dimensional Quantum Physics and Department of Physics, Tsinghua University, Beijing 100084, China}

\author{Eryin Wang}
\affiliation{State Key Laboratory of Low Dimensional Quantum Physics and Department of Physics, Tsinghua University, Beijing 100084, China}

\author{Mingzhe Yan}
\affiliation{State Key Laboratory of Low Dimensional Quantum Physics and Department of Physics, Tsinghua University, Beijing 100084, China}

\author{Huaqing Huang}
\affiliation{State Key Laboratory of Low Dimensional Quantum Physics and Department of Physics, Tsinghua University, Beijing 100084, China}

\author{Hongyun Zhang}
\affiliation{State Key Laboratory of Low Dimensional Quantum Physics and Department of Physics, Tsinghua University, Beijing 100084, China}

\author{Zhilin Xu}
\affiliation{State Key Laboratory of Low Dimensional Quantum Physics and Department of Physics, Tsinghua University, Beijing 100084, China}

\author{Jonathan Denlinger}
\affiliation{Advanced Light Source, Lawrence Berkeley National Laboratory, Berkeley, California 94720, USA}

\author{Alexei Fedorov}
\affiliation{Advanced Light Source, Lawrence Berkeley National Laboratory, Berkeley, California 94720, USA}

\author{Haitao Yang}
\affiliation{State Key Laboratory of Low Dimensional Quantum Physics and Department of Physics, Tsinghua University, Beijing 100084, China}
\affiliation{Tsinghua-Foxconn Nanotechnology Research Center, Tsinghua University, Beijing 100084, China}

\author{Wenhui Duan}
\affiliation{State Key Laboratory of Low Dimensional Quantum Physics and Department of Physics, Tsinghua University, Beijing 100084, China}
\affiliation{Collaborative Innovation Center of Quantum Matter, Beijing, China}

\author{Hong Yao}
\affiliation{Institute for Advanced Study, Tsinghua University, Beijing 100084, China}
\affiliation{Collaborative Innovation Center of Quantum Matter, Beijing, China}

\author{Yang Wu}
\altaffiliation{Correspondence should be sent to wuyangthu@mail.tsinghua.edu.cn, xc@mail.tsinghua.edu.cn and syzhou@mail.tsinghua.edu.cn}
\affiliation{Tsinghua-Foxconn Nanotechnology Research Center, Tsinghua University, Beijing 100084, China}

\author{Shoushan Fan}
\affiliation{State Key Laboratory of Low Dimensional Quantum Physics and Department of Physics, Tsinghua University, Beijing 100084, China}
\affiliation{Tsinghua-Foxconn Nanotechnology Research Center, Tsinghua University, Beijing 100084, China}
\affiliation{Collaborative Innovation Center of Quantum Matter, Beijing, China}

\author{Haijun Zhang}
\affiliation{National Laboratory of Solid State Microstructures and School of Physics,  Nanjing University, Nanjing 210093, China}
\affiliation{Collaborative Innovation Center of Advanced Microstructures, Nanjing, China}

\author{Xi Chen}
\altaffiliation{Correspondence should be sent to wuyangthu@mail.tsinghua.edu.cn, xc@mail.tsinghua.edu.cn and syzhou@mail.tsinghua.edu.cn}
\affiliation{State Key Laboratory of Low Dimensional Quantum Physics and Department of Physics, Tsinghua University, Beijing 100084, China}
\affiliation{Collaborative Innovation Center of Quantum Matter, Beijing, China}

\author{Shuyun Zhou}
\altaffiliation{Correspondence should be sent to wuyangthu@mail.tsinghua.edu.cn, xc@mail.tsinghua.edu.cn and syzhou@mail.tsinghua.edu.cn}
\affiliation{State Key Laboratory of Low Dimensional Quantum Physics and Department of Physics, Tsinghua University, Beijing 100084, China}
\affiliation{Collaborative Innovation Center of Quantum Matter, Beijing, China}

\date{\today}

\maketitle

{\bf
Weyl semimetal is a new quantum state of matter \cite{nielsen1983, WanXGPRB11, burkov2011, FangZPRL11, RanYPRB11, hosur2013,zhang2014a,vanderbilt2014, hirayama2015, DaiXTaAs, HuangSM, Ruan-HgTe15} hosting the condensed matter physics counterpart of relativistic Weyl fermion \cite{Weyl} originally introduced in high energy physics.  The Weyl semimetal realized in the TaAs class features multiple Fermi arcs arising from topological surface states \cite{DaiXTaAs, HuangSM, HasanSci15, DingHTaAsPRX, ChenTaAs} and exhibits novel quantum phenomena, e.g., chiral anomaly induced negative magnetoresistance \cite{son2013,TaAsMR,zhang2015a} and possibly emergent supersymmetry \cite{Jian2015}. Recently it was proposed theoretically that a new type (type-II) of Weyl fermion \cite{BernevigNat15}, which does not have counterpart in high energy physics due to the breaking of Lorentz invariance, can emerge as topologically-protected touching between electron and hole pockets.
Here, we report direct experimental evidence of topological Fermi arcs in the predicted type-II Weyl semimetal MoTe$_2$ \cite{YanBHprb,BernevigMoTe2,MoTe2qpi}.  The topological surface states are confirmed by directly observing the surface states using bulk- and surface-sensitive angle-resolved photoemission spectroscopy (ARPES), and the quasi-particle interference (QPI) pattern between the putative topological Fermi arcs in scanning tunneling microscopy (STM). Our work establishes MoTe$_2$ as the first experimental realization of type-II Weyl semimetal, and opens up new opportunities for probing novel phenomena such as exotic magneto-transport \cite{BernevigNat15} in type-II Weyl semimetals.
}

\begin{figure*}
\includegraphics[width=16 cm] {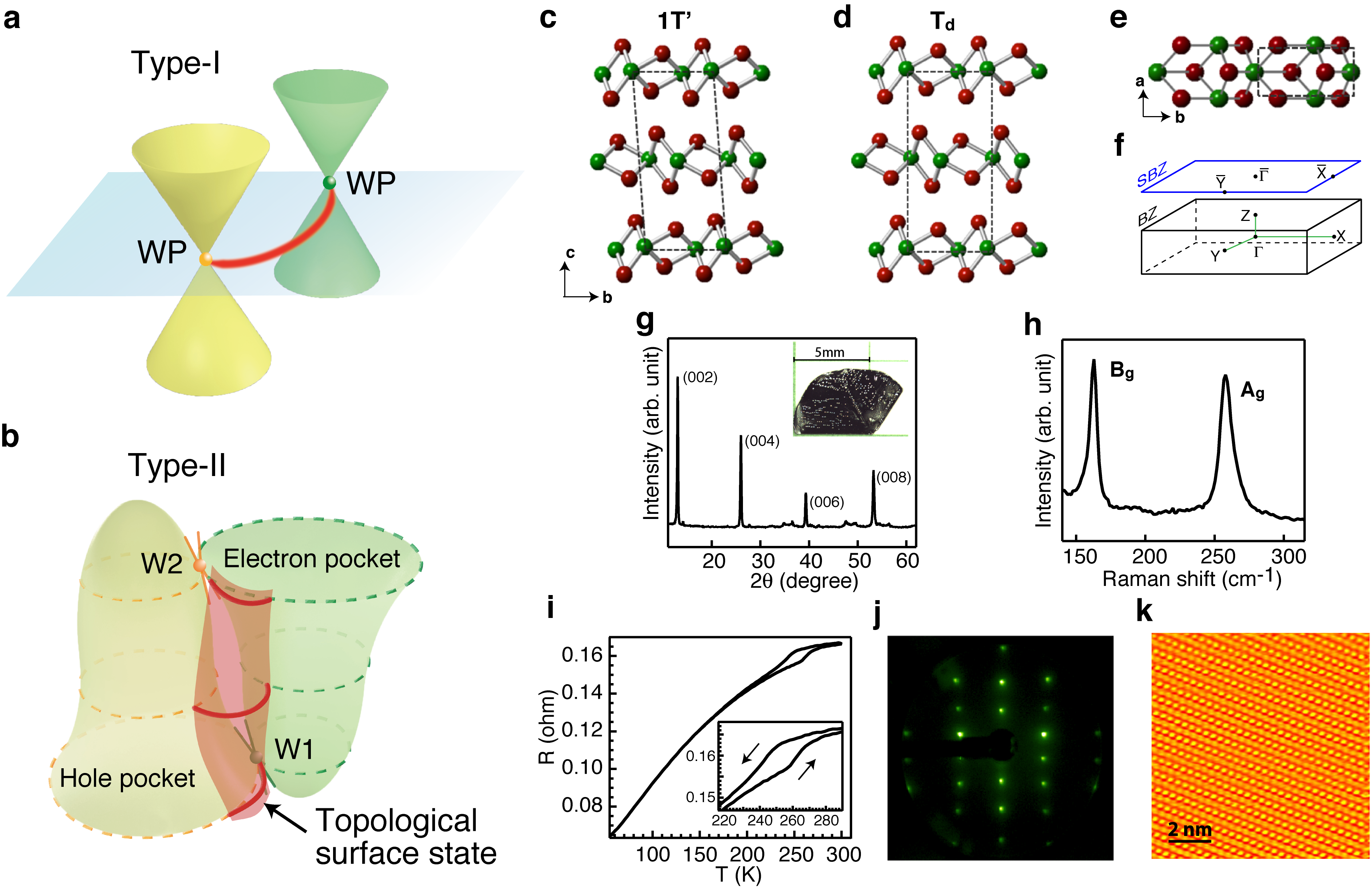}
\caption{{\bf Characterization of type-II Weyl semimetal MoTe$_2$.} (a)  Dispersions for type-I Weyl fermion near $E_F$. The Weyl points (WP) are labeled by yellow and green dots. (b)  Type-II Weyl semimetal with electron and hole pockets touching at two different energies. (c-d)  Crystal structures of MoTe$_2$ in the 1T$^\prime$ (c)  and T$_d$ (d) phases. Green balls are Mo atoms and red balls are Te atoms. (e) The in-plane crystal structure. (f) Bulk and projected surface Brillouin zone. (g) XRD of MoTe$_2$ measured at room temperature (1T$^\prime$ phase). The inset shows a picture of the few mm size single crystal. (h) Raman spectrum measured at room temperature. (i) Transport measurement shows a first order phase transition between the 1T$^\prime$ phase and the T$_d$ phase. (j) LEED pattern taken in the T$_d$ phase at beam energy of 180 eV. (k) The STM topography (bias -50 mV, tunneling current 0.05 nA) taken at 4.2 K.  }
\label{Characterization}
\end{figure*}

In the Brillouin zone of a type-I Weyl semimetal, the linearly dispersing and non-degenerate bands cross each other at the Weyl points (Fig.~\ref{Characterization}(a)). These band-topology protected Weyl points  can only be created or annihilated in pairs according to the no-go theorem \cite{nielsen1983}.  When projected onto the surface, the Weyl points are connected by the topologically protected Fermi arcs (Fig.~\ref{Characterization}(a)) \cite{WanXGPRB11}. In contrast to the type-I Weyl fermions in TaAs class or compressively-strained HgTe \cite{Ruan-HgTe15} which have point-like Fermi surface, the type-II Weyl fermions emerge at the boundary between electron and hole pockets when the cones are tilted significantly (Fig.~\ref{Characterization}(b)), and there are finite density of states at  the Fermi energy $E_F$.  The distinction between the Fermi surfaces of these two types of Weyl semimetals is expected to lead to different physical properties and response to magnetic fields \cite{BernevigNat15}.

Type-II Weyl fermion has been predicted in the orthorhombic T$_d$ phase of WTe$_2$ \cite{BernevigNat15}, which breaks the inversion symmetry and shows unusual transport properties \cite{WTe2MR}. However, the small momentum separation of the Weyl points (0.7$\%$ of the Brillouin zone) and the extremely small size of the arcs \cite{BernevigNat15} make it exceptionally challenging to resolve the topological Fermi arcs in WTe$_2$ by ARPES.  A promising solution is provided by the prediction that the topological Fermi arcs can be significantly enlarged in MoTe$_2$ \cite{YanBHprb, BernevigMoTe2}  or Mo$_x$W$_{1-x}$Te$_2$ \cite{HasanMoWTe2}. Among these candidate materials, MoTe$_2$ is particularly interesting because of the reported superconductivity \cite{YanBHSC} and the predicted topological phase transition induced by temperature or strain \cite{YanBHprb}. Although the electronic structures of WTe$_2$ \cite{VallaWTe2, KaminskiWTe2, FengDLWTe2} and Mo$_x$W$_{1-x}$Te$_2$ \cite{HasanTRARPES} have been experimentally studied, so far there is no conclusive evidence on the existence of topological Fermi arcs.  Here by combining two complementary surface sensitive probes -  ARPES and STM, we provide direct experimental evidence of the topological Fermi arcs at the boundary between electron and hole pockets in the T$_d$ phase of MoTe$_2$, establishing it as a type-II Weyl semimetal.

MoTe$_2$ is polymorphic with three different structures: hexagonal ($\alpha$-phase, or 2H phase), monoclinic ($\beta$-phase, or 1T$^\prime$ phase) and orthorhombic ($\gamma$-phase, or T$_d$ phase). The 1T$^\prime$ phase has a distorted CdI$_2$ structure (Fig.~\ref{Characterization}(c)) that crystalizes in the centrosymmetric space group P2$_1$/m. The Mo atoms are coordinated by six Te atoms but shifted from the center of the Te octahedra, resulting in the zigzag chains along the b axis. The bonding between the shifted Mo atoms corrugates the Te sheets and distorts the Te octahedra \cite{Clarke, Manolikas79}, causing the c axis to incline at an angle of $\sim$ 93.9$^\circ$ \cite{Clarke}. A temperature induced phase transition from the high temperature 1T$^\prime$  to the low temperature T$_d$  phase has been reported between 240 K to 260 K \cite{Clarke}. The T$_d$  phase (Fig.~\ref{Characterization}(d)) shares the same in-plane crystal structure (Fig.~\ref{Characterization}(e)) as 1T$^\prime$ phase but has a vertical (90$^\circ$) stacking and belongs to the non-centrosymmetric space group Pmn2$_1$.  Weyl fermions are only possible in the T$_d$ phase where the inversion symmetry is broken. The Brillouin zone of T$_d$  phase is shown in Fig.~\ref{Characterization}(f).

\begin{figure*}
\includegraphics[width=16 cm] {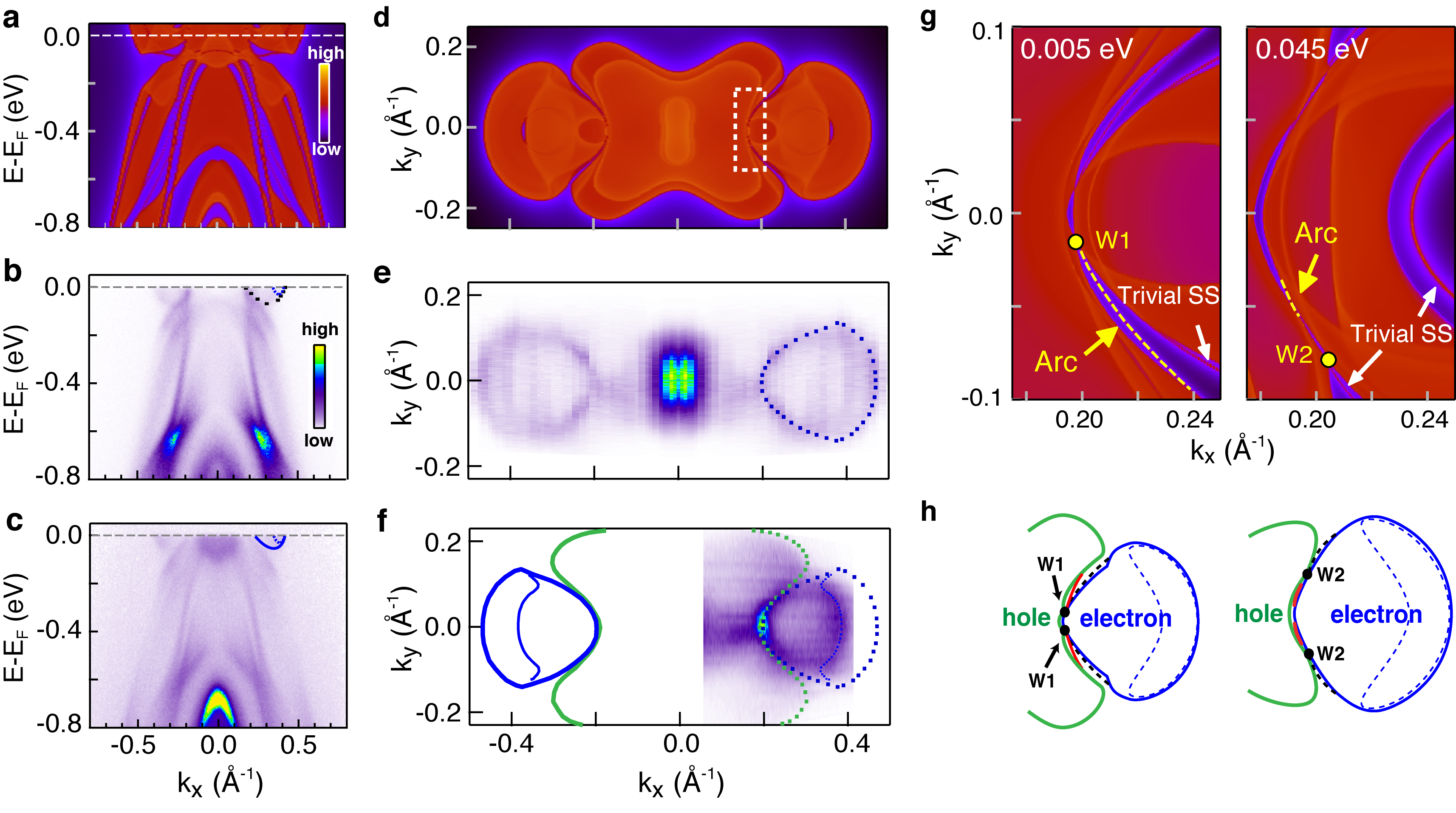}
\caption{{\bf Band structure of MoTe$_2$.} (a) Calculated dispersion along the $\bar{X}$-$\bar{\Gamma}$-$\bar{X}$ direction. (b,c) Measured dispersions along the $\bar{X}$-$\bar{\Gamma}$-$\bar{X}$ direction with horizontal (p) and vertical (s) polarizations at photon energy of 32.5 eV.  A comparison of calculated band structure with experimental dispersions show that the sample is slightly hole doped, and the calculated energy needs to be shifted by -0.02 eV.  (d) Calculated spectral function at $E_F$.   (e,f) Intensity maps measured at $E_F$ with p-polarization using 6.3 eV laser source with light polarizations perpendicular to the b- and a- axis  respectively. Map in (e) was obtained by symmetrizing the data taken at positive k$_y$ values. The electron and hole pockets are highlighted by blue and green color. (g) Calculated spectral function at 0.005 eV and 0.045 eV. The range of momentum is marked by dashed rectangle in (d). The Weyl points and topological Fermi arcs are highlighted.   (h) Schematic showing the various electronic structure at  the Weyl points W1 and W2.}
\label{Arpes}
\end{figure*}

Figure \ref{Characterization}(g) shows the X-ray diffraction (XRD) intensity of the high quality MoTe$_2$ single crystal at room temperature (1T$^\prime$ phase). The Raman spectrum in Fig.~\ref{Characterization}(h) shows B$_g$ and A$_g$ vibrational modes at $\sim$ 160 and 260 cm$^{-1}$ respectively, consistent with other report \cite{LeeMoTe2NatPhys}.  The resistance measurement (Fig.~\ref{Characterization}(i)) confirms the first order phase transition between the T$_d$ and 1T$^\prime$ phases at $\sim$ 260 K, in agreement with previous results \cite{Clarke}. The high crystallinity of the samples is revealed by the sharp diffraction spots (Fig.~\ref{Characterization}(j)) in the low energy electron diffraction (LEED) pattern measured on a freshly cleaved sample in the T$_d$ phase. The atomically resolved STM topography in Fig.~\ref{Characterization}(k) further confirms the high quality of the MoTe$_2$ crystal. The cleaved surface is terminated by Te atoms. The image shows a rectangular lattice with the lattice constants of a = 3.5 \AA, b = 6.3 \AA. The center and corner atoms of a rectangular unit are different in height and exhibit distinct contrast. The d$I$/d$V$ spectrum on the surface is shown in the supplementary information.

Figure~\ref{Arpes}(a-c) compares the electronic structure of MoTe$_2$ in the T$_d$ phase measured by ARPES with band structure calculation along the a-axis ($\bar{X}$-$\bar{\Gamma}$-$\bar{X}$) direction. The bands with significant k$_z$ dispersion overlap to form continuously filled contours, while those with strong surface state characteristics show up as sharp features in the intensity maps. The ARPES spectral intensity is affected by the dipole matrix elements and thereby depends on both the electron wave function and light polarization. To resolve the dispersions of multiple pockets, we use UV light with both horizontal (p) and vertical (s) polarizations. The measured dispersions (Fig.~\ref{Arpes}(b,c)) are in good agreement with the first-principles calculations (Fig.~\ref{Arpes}(a)). The trivial surface states (marked by the black broken curve) together with the smaller electron pocket (blue broken curve) are better resolved with the p-polarization light (Fig.~\ref{Arpes}(b)), while the s-polarization light  (Fig.~\ref{Arpes}(c)) clearly resolves both  bulk electron pockets (blue solid and dotted curves) and the pocket surrounding the $\Gamma$ point.  In the calculated spectral function (Fig.~\ref{Characterization}(d)), the spectral weight of the electron pockets forms bell-like shapes on both sides away from the $\Gamma$ point and part of the bowtie-shaped outer contour around the  $\Gamma$ point is contributed by the hole pockets at $E_F$. These bulk states are better observed with bulk sensitive laser source at 6.3 eV  (penetration depth of $\approx$ 30 \AA) in ARPES. Figure \ref{Arpes}(e,f) shows the measured Fermi surface maps with light polarizations perpendicular to the b- and a- axis respectively. The bulk electron pockets are clearly observed in Fig.~\ref{Arpes}(e) and have an overall uniform intensity contour  (blue broken curve), while the bowtie-shaped hole pocket is more clearly observed in Fig.~\ref{Arpes}(f) (green curve). 

\begin{figure*}
\includegraphics[width=15 cm] {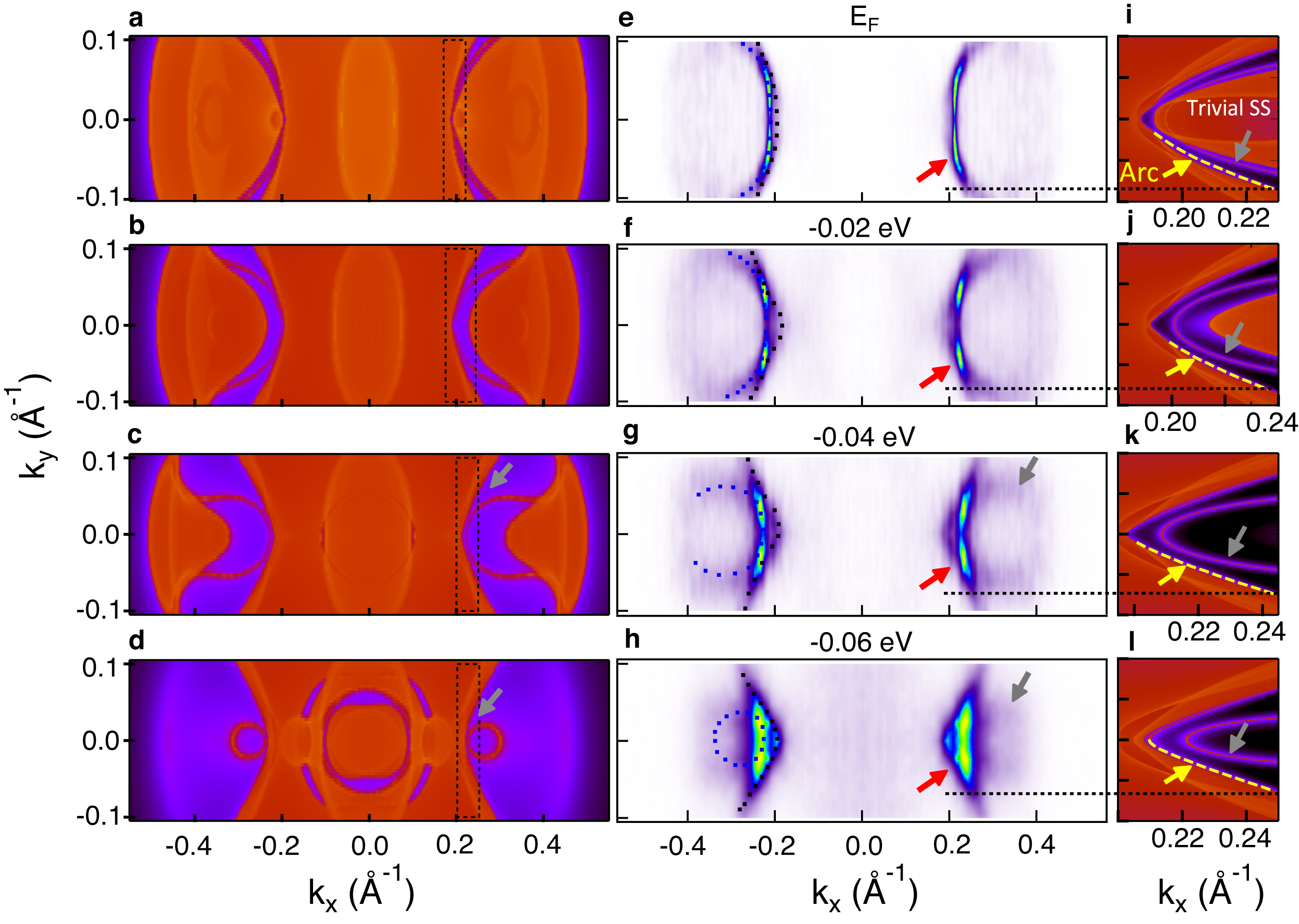}
\caption{{\bf Observation of topological Fermi arcs in the T$_d$ phase of MoTe$_2$.} (a-d) Calculated spectral intensity maps (shifted by -0.02 eV to account for the slight hole doping). The rectangles mark the regions which are enlarged in (i-l). (e-h) ARPES intensity maps at energies from E$_F$ to -0.06 eV. The maps were symmetrized with respect to k$_y$ and k$_x$. Red and gray arrows point to the topological and trivial surface states, respectively.  (i-l) Zoom-in of rectangular regions in (a-d) to show the arcs from the topological surface states. The arcs at negative k$_y$ are highlighted by yellow broken curves. Yellow arrows point to topological surface states and gray arrows point to trivial surface states. Black dotted lines are guides to the eye for the termination points of the arcs from ARPES measurement and calculation.}
\label{Maps}
\end{figure*}

According to  band structure calculation (Fig.~\ref{Arpes}(g)), the above observed electron and hole pockets  touch each other at eight Weyl points with energies of $\approx$ 0.005 eV (W1) and $\approx$ 0.045 eV (W2), respectively.  Topological Fermi arcs (highlighted by yellow curves in Fig.~\ref{Arpes}(g)) are expected to emerge between the Weyl points with opposite chiralities \cite{BernevigNat15, YanBHprb, BernevigMoTe2}. At the energy of W2, part of the arcs are shadowed by the pockets and only a small portion is observed.
In addition to the topological surface states, there are also trivial surface states (pointed by white arrows). Theoretical calculation also shows that in the centrosymmetric 1T$^\prime$ phase, the electron and hole pockets have no touching points, and only the trivial surface states remain (see Fig.~S1 in supplementary information). The disappearance of the Fermi arcs in the 1T$^\prime$ phase further confirms their origin from the Weyl semi-metallic state. The characteristic electronic structure of T$_d$ phase MoTe$_2$ is schematically summarized in Fig.~\ref{Arpes}(h) with the  energies of the Weyl points as examples. 

\begin{figure*}
\includegraphics[width=13 cm] {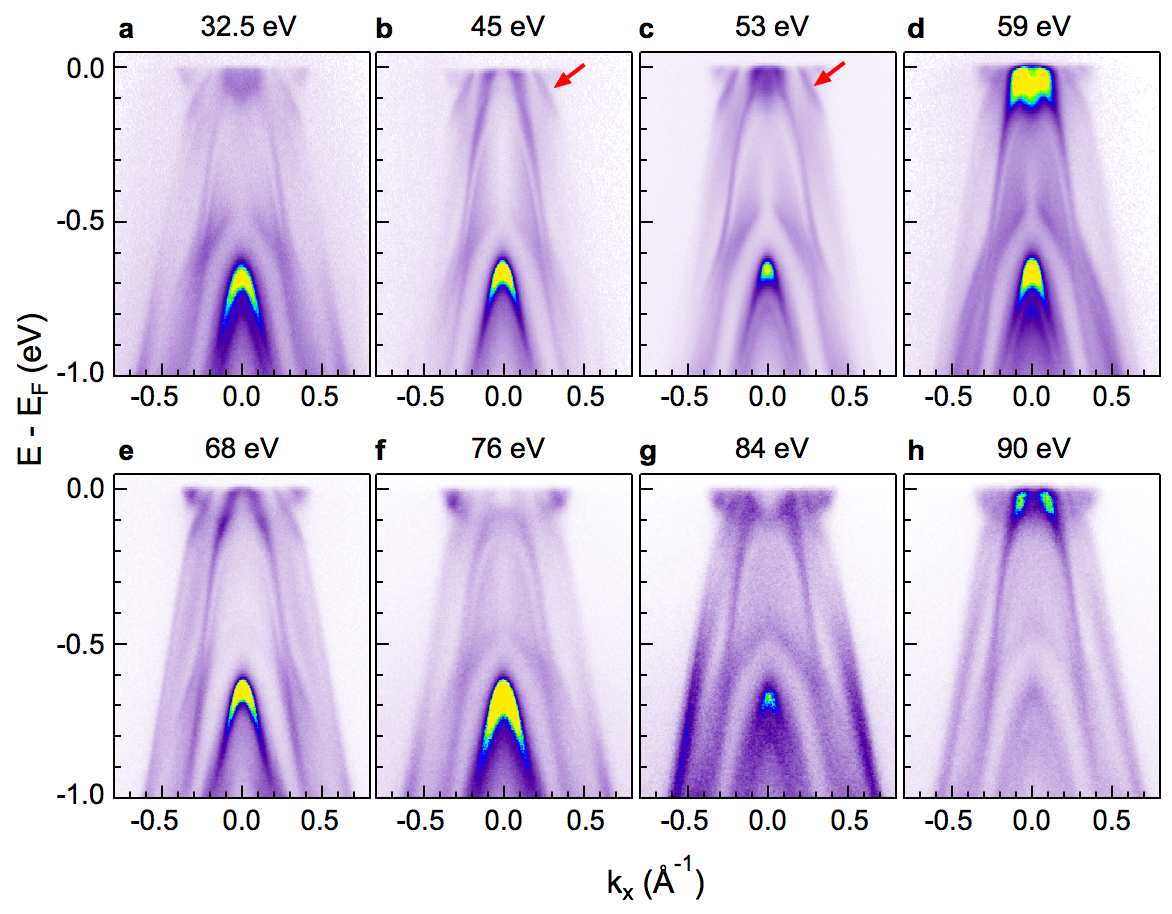}
\caption{{\bf Bulk states versus surface states using ARPES.} (a-h) Intensity maps measured along the $\bar{X}$-$\bar{\Gamma}$-$\bar{X}$ direction at selected photon energies from 32.5 eV to 90 eV. Due to the finite analyzer slit width, the measured dispersion is averaged over a finite k$_y$ momentum window (e.g. $\pm$ 0.026 $\AA$ at 53 eV photon energy) and the measured dispersion covers contribution from the topological surface states which start from k$_y$ $\approx$ 0.014 $\AA$ (see Fig.~2 in supplementary information).}
\label{hvdep}
\end{figure*}

Since both the topological and the trivial surface states are squeezed in the narrow gap between the electron and hole pockets (Fig.~\ref{Maps}(a-d)), resolving the different features in ARPES measurement is the most challenging aspect to correctly identify the topological Fermi arcs. We search for the topological Fermi arcs in  ARPES intensity maps with surface sensitive UV source  (penetration depth of a few \AA). The intensity contribution from bulk bands is largely suppressed by using selected specific surface sensitive photon energy with different polarizations, and the surface states in-between the bulk electron and hole pockets can thus become more accessible experimentally.

Figure~\ref{Maps}(e-h) shows the high resolution ARPES intensity maps taken at 32.5 eV photon energy.  The arcs (pointed by red arrows) are clearly observed.  At E$_F$ (panels (a), (e) and (i)) and -0.02 eV (panel (b), (f) and (j)), the arcs and the trivial surface states are not well separated. However, as the electron pocket shrinks with decreasing energy, the separation between the topological Fermi arcs (red arrow in panel (g) and yellow arrow in panel (k)) and the trivial surface states (indicated by gray arrow in panels (c), (g) and(h)) becomes more pronounced. At -0.06 eV where the electron pocket completely disappears (panel (d)), the trivial surface states form a loop (panels (d) and (h)) and are clearly separated from the hole pocket.  The evolution of the topological and trivial surface states in ARPES measurement is in good agreement with that from the band structure calculation.  Furthermore, a comparison with the zoom-in calculated maps  shows that the termination points of the observed arcs (panel (e-h)) line up with those of the calculated ones (yellow broken curves in panels (i-l)), explicitly supporting the presence of topological Fermi arcs.

\begin{figure*}
\includegraphics[width=15.8 cm] {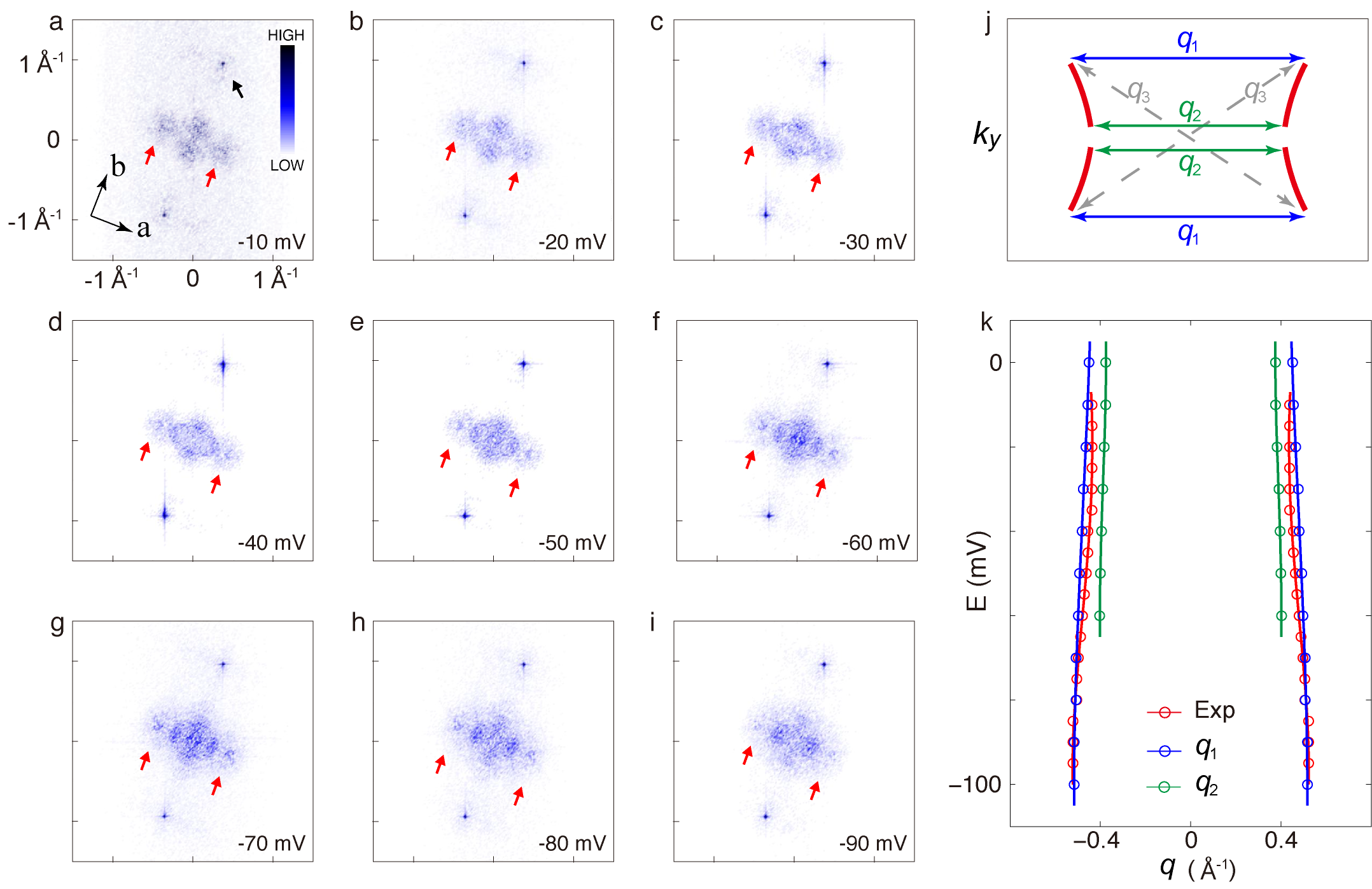}
\caption{{\bf Quasi-particle interference pattern.} (a-i) The FFT power spectra of the dI/dV maps of a region shown in supplementary information. The current was set at 0.1 nA. Each map has 256 $\times$ 256 pixels. The temperature is 4.2 K. The features produced by Fermi arc are indicated by red arrows.  The origin of the additional features in QPI needs further investigation. The black arrow points to a Bragg peak. (j) The extremal pairs owing to a pair of topological Fermi arcs.  (k) Dispersions  extracted from QPI and comparison with the calculated $q_1$ and $q_2$. The topological Fermi arcs at positive and negative k$_y$ values merge beyond -0.06 eV, and therefore q$_2$ is plotted only down to -0.06 eV.
}
\label{qpi}
\end{figure*}

The observed topological Fermi arcs reside on the two-dimensional (2D) crystal surface. We performed more experimental studies, including variable incident photon energy measurement and quasi-particle interference in real space, to support the surface nature of the observed electronic feature.  Bulk states with different k$_z$ values selectively respond to different incident photon energy, which helps to separate the contributions from bulk and surface states. Figure \ref{hvdep}(a-h) shows ARPES data measured  along the $\bar{X}$-$\bar{\Gamma}$-$\bar{X}$ direction with photon energies from 32.5 eV to 90 eV.  The dispersions near the $\bar{\Gamma}$ point change significantly with incident photon energy, suggesting that they are from bulk states.  In contrast, the previously identified surface band (between E$_F$ and -0.1 eV  and pointed by red arrows in panels (b) and (c)) appears at the same position with different photon energies.  Consistently, this surface band is most clearly observed at 45 eV and 53 eV,  where the penetration depth of photon reaches the minimum.

The complementary surface sensitive probe-STM provides another independent experimental evidence to support the surface nature of the arcs. Universal signatures of topological Fermi arcs in QPI on the surface of Weyl semimetals have been theoretically established by Ref.~\onlinecite{MoTe2qpi}. Various defects on the surface elastically scatter the electrons and induce the QPI pattern. In the surface Brillouin zone, the extremal pairs of  $\vec{k}_i$ and $\vec{k}_f$ on a 2D constant energy contour, where $\vec{k}_i$ and $\vec{k}_f$ are the initial and final wave vectors, contribute dominantly to the spatial interference pattern of the local electron density of states (LDOS) \cite{ChenQPI}. The spatial variation of LDOS at a certain energy is the sum of the contributions from all the extremal pairs on the constant energy contour and measured by the differential conductance (dI/dV) mapping with spatial resolution. The features in the Fourier transform of dI/dV mapping correspond to the scattering vector $\vec{Q}=\vec{k}_f-\vec{k}_i$ of the extremal pairs. QPI is more sensitive to the surface states or states with small $k_z$ dependence than to the bulk ones  with strong $k_z$ dependence since the latter cannot host the ``extreme pairs''. In this sense, QPI is advantageous in studying type-II Weyl semimetal MoTe$_2$, where the topological Fermi arcs and the projected bulk pockets are very close in energy.

Figure~\ref{qpi}(a-i) displays the fast Fourier transform (FFT) of the dI/dV maps between -10 mV and -90 mV. For a pair of topological Fermi arcs, three scattering wave vectors (Fig.~5(j)), labeled $q_1$, $q_2$, and $q_3$, might be expected to appear in QPI. Among them, $q_3$ is forbidden due to the requirement of the time-reversal symmetry in the system. Similar forbidden scattering was also experimentally observed in the surface states of topological insulators with time-reversal symmetry \cite{TIQPI}.  The scattering wave vectors should generate visible features centered between $q_1$ and $q_2$ and along the $\Gamma$-X direction (Fig.~\ref{qpi}(j)). Such features are  clearly resolved and indicated by red arrows in FFT. The existence of such pattern beyond the band bottom of the trivial surface states (-60 mV) excludes the possibility of trivial surface states as the origin. Moreover, the dispersions extracted from the energy dependent scattering wave vector (panel (k)) is in very good agreement with the $q_1$ and $q_2$ extracted from band structure calculation, providing another independent and strong evidence for the existence of topological surface states. By combining two complementary surface sensitive experimental probes - STM, ARPES - with theoretical calculations, we provide direct and strong experimental evidence for the existence of the topological surface states, establishing it as a type-II Wey semimetal.

{\it Note added.}  During revision of this manuscript for resubmission, we became aware of related work by L. Huang {\it et al.} \cite{Kaminski} and by S-Y Xu.{\it et al.} \cite{HasanLAG}.

{\bf Methods}

{\bf Sample growth.} High quality $\beta$-MoTe$_2$ single crystals  were grown by chemical vapor transport using polycrystalline MoTe$_2$ as precursors. Polycrystalline MoTe$_2$ was synthesized by directly heating the stoichiometric mixture of high-purity Mo foil (99.95$\%$, Alfa Aesar) and Te ingot (99.99$\%$, Alfa Aesar)  at 1073 K in a vacuum-sealed silica ampoule for 3 days. The as-grown MoTe$_2$ was then recrystallized by the chemical vapor transport method using powder TeCl$_4$ (99$\%$, Aladdin) as transporting agent with a concentration of $\le$ 2.7 mg/mL. Material transport occurred in a sealed silica ampoule in a tube furnace for 3 days. After the reaction, the ampoule was immediately quenched in cold water to obtain large size $\beta$-MoTe$_2$ single crystals.

{\bf ARPES measurement.} Bulk sensitive laser-ARPES measurements have been performed in the home laboratory at Tsinghua University with four harmonic generation light source. Surface sensitive ARPES measurements have been performed at BL.4.0.1 and BL.12.0.1 of the Advanced Light Source using photon energies from 30.5 eV to 90 eV. The overall experimental energy resolution at 32.5 eV is better than 18 meV. The samples were cleaved and measured at 10-20 K in the T$_d$ phase.

{\bf STM measurement.} STM experiments were conducted on a Unisoku ultrahigh vacuum low temperature (down to 4.2 K) system equipped with {\it in situ} cleaving stage. The MoTe$_2$ single crystals were cleaved in ultrahigh vacuum (5$\times$10$^{-11}$ Torr) at room temperature and then transferred to STM to perform measurement at 4.2 K with a PtIr tip. QPI maps and dI/dV spectra were acquired using a lock-in amplifier at frequency of 913 Hz.

{\bf First-principles calculations.}

The {\it ab-initio} calculations are carried out in the framework of the Perdew-Burke-Ernzerhof-type generalized gradient approximation of the density functional theory through employing the Vienna Ab initio simulation package (VASP) \cite{kresse1996} with the projected augmented wave (PAW) method. The kinetic energy cutoff is fixed to 400 eV, and the {\bf k}-point mesh is taken as 12$\times$10$\times$6 for the bulk calculations. The spin-orbit coupling effect is self-consistently included. The lattice constants are taken from experiments \cite{YanBHprb}, but the atoms in the unit cell are fully relaxed with the force cutoff 0.01eV/\AA. Maximally localized Wannier functions are employed to obtain the  {\it ab-initio} tight-binding model of semi-infinite systems with the (001) surface as the  boundary to exhibit surface states and topological Fermi arcs. An iterative method  is used to obtain the surface Green's function of the semi-infinite system.

{\bf Acknowledgements}
This work is supported by the National Natural Science Foundation of China (grant No.~11274191, 11334006), Ministry of Science and Technology of China (No.~2015CB92100, 2016YFA0301004 and 2012CB932301) and Tsinghua University Initiative Scientific Research Program (No.~2012Z02285). The Advanced Light Source is supported by the Director, Office of Science, Office of Basic Energy Sciences, of the US Department of Energy under Contract No.~DE-AC02-05CH11231.

{\bf Author Contributions}
S.Z., X.C. and Y.W. conceived the research project. K.D. and K.Z. grew and characterized the samples under supervision of Y.W.. K.D., G.W, K.Z., S.D., E.W., M.Y., and H.Y.Z. performed the ARPES measurements and analyzed the ARPES data. J.D. and A. F.provided support for the ARPES experiments.  P.D. and Z.X. performed the STM measurements. H.J.Z. performed the first principles calculations presented in the manuscript. H.H and W.D. repeated the calculation. K.D., H.Y., Y.W., X.C. and S.Z. wrote the manuscript, and all authors commented on the manuscript.

{\bf Competing financial interests}
The authors declare no competing financial interests.

\clearpage

{\bf SUPPLEMENTARY INFORMATION}

\section{A comparison of the electronic structure for T$_d$ phase and T$^\prime$ phase from band structure calculation}

To confirm the Weyl points and identify the topological surface states in the T$_d$ phase, we show in Fig.~1 a comparison of the calculated electronic structure of the inversion asymmetric T$_d$ phase and the inversion symmetric 1T$^\prime$ phase. In the T$_d$ phase, the electron and hole pockets touch at two sets of Weyl points W2 ($\approx$ 45 meV) and W1 ($\approx$ 5 meV) respectively (Fig.~1(c,d)). Fermi arcs can be clearly identified (highlighted by yellow broken line). In addition, there are also trivial surface states which are connected to the electron pockets. In the 1T$^\prime$ phase, there is no crossing point between the electron and hole pockets (i.e. no Weyl points). Consequently Fermi arcs disappear (Fig.~1(g,h), leaving only the trivial surface states. This confirms that the Weyl points and Fermi arcs arise only in the T$_d$ phase when the inversion symmetry is broken, providing direct evidence for the Weyl semimetallic state in the T$_d$ phase of MoTe$_2$ from band structure calculation. 

\begin{figure*}
\includegraphics[width=16.8 cm] {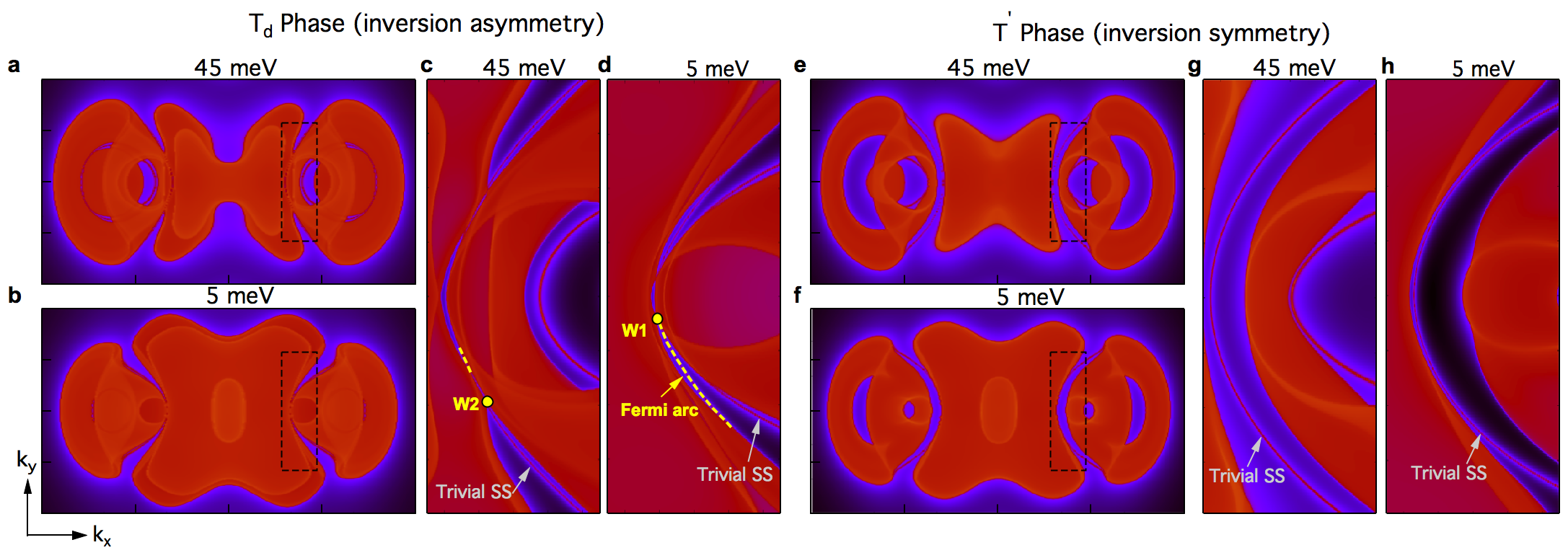}
\caption*{{\bf Fig.S1: Confirmation of topological Fermi arcs in the T$_d$ phase from numerical calculation.} (a,b) Projected intensity maps at the Weyl point energies for W2 and W1. (c,d) Zoom-in of the regions marked by rectangles in (a,b) to see the Fermi arcs. The  topological Fermi arcs are marked by yellow arrow and trivial surface states are pointed by gray arrows. (e,f) Projected intensity maps at the same energies as (a,b) for the 1T$^\prime$ phase. The electron and hole pockets are separated. (g,h) Zoom-in of the regions marked by rectangles. Only trivial surface states are observed.}
\end{figure*}

\begin{figure*}
\includegraphics[width=16.8 cm] {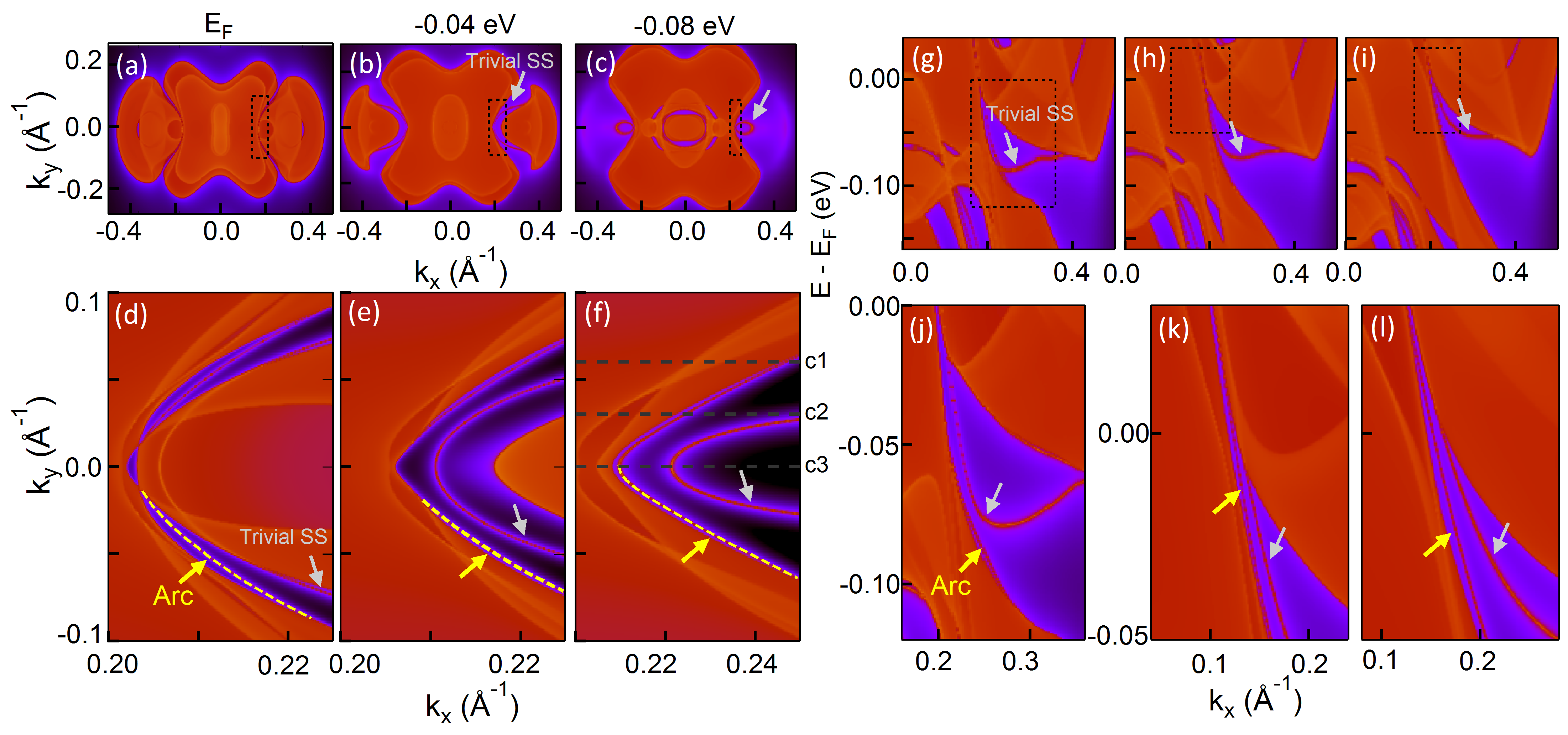}
\caption*{{\bf Fig.S2: Topological surface states in the T$_d$ phase from band structure calculations.} (a-c) Projected intensity maps at E$_F$, -0.04 and -0.08 eV. (d-f) Zoom-in of the regions marked by rectangles in (a-c). Topological Fermi arcs are pointed by yellow arrows and trivial surface states are pointed by gray arrows. (g-i) Projected dispersions parallel to the $\bar{X}$-$\bar{\Gamma}$-$\bar{X}$ direction at selected k$_y$ positions marked by horizontal broken lines in (f). (j-l) Zoom-in of regions marked by rectangles in (g-i) to show the trivial surface states and topological surface states. }
\end{figure*}

To further identify the regions where the topological surface states can be observed, we show in Fig.~3 a systematic calculation of the intensity maps and projected dispersions parallel to $\bar{X}$-$\bar{\Gamma}$-$\bar{X}$ direction.

\section{STM measurements}

\begin{figure*}
\includegraphics[width=8 cm] {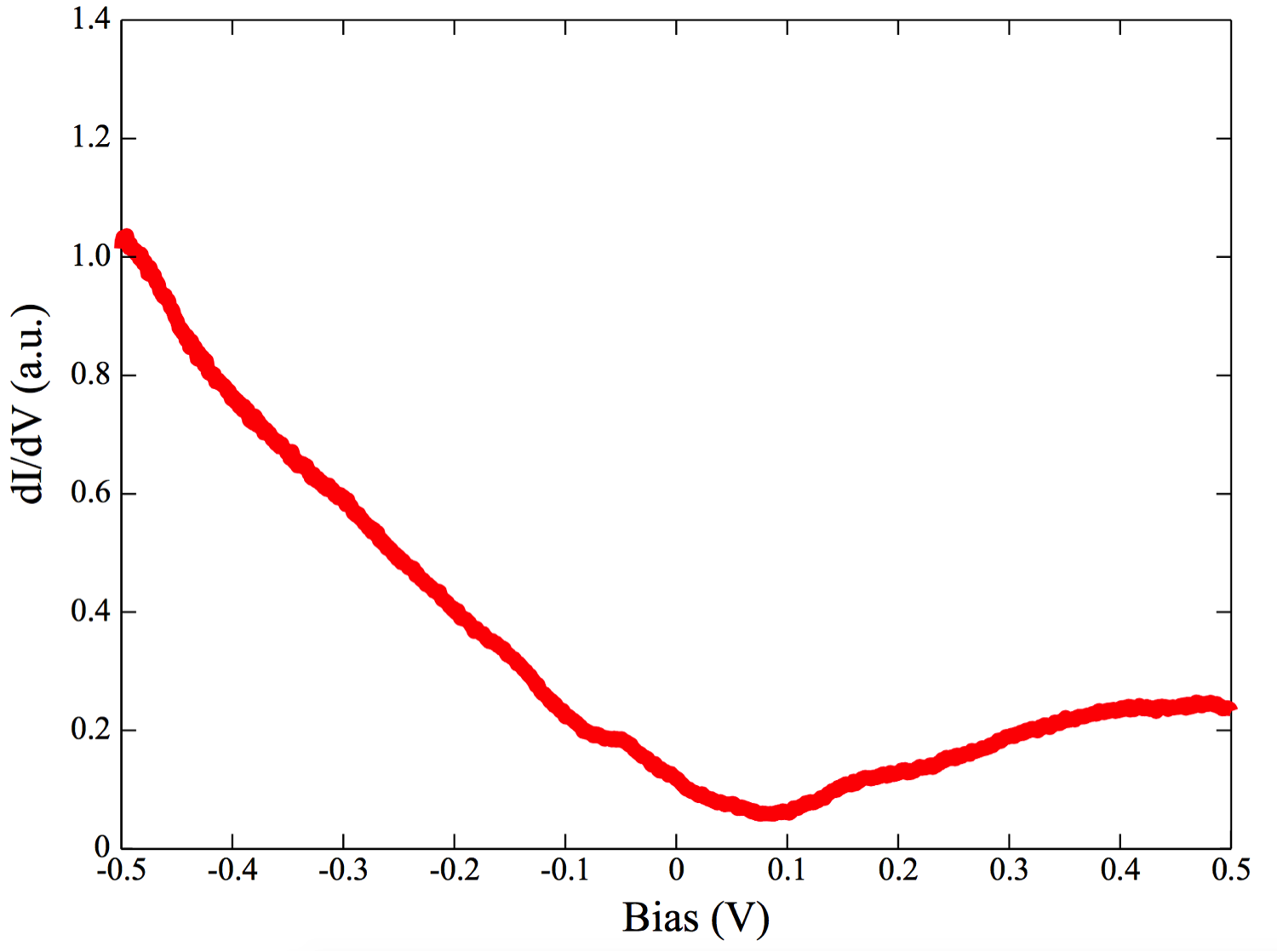}
\caption*{{\bf Fig.S3: dI/dV spectrum.} Temperature: 4.2 K. Setpoint: 500 mV, 0.1 nA. Lockin oscillation amplitude: 5 mV. The dI/dV spectrum measures the local density of states of electrons. The minimum at 0.07 V corresponds to the top of hole pocket.}
\end{figure*}

\begin{figure*}
\includegraphics[width=8 cm] {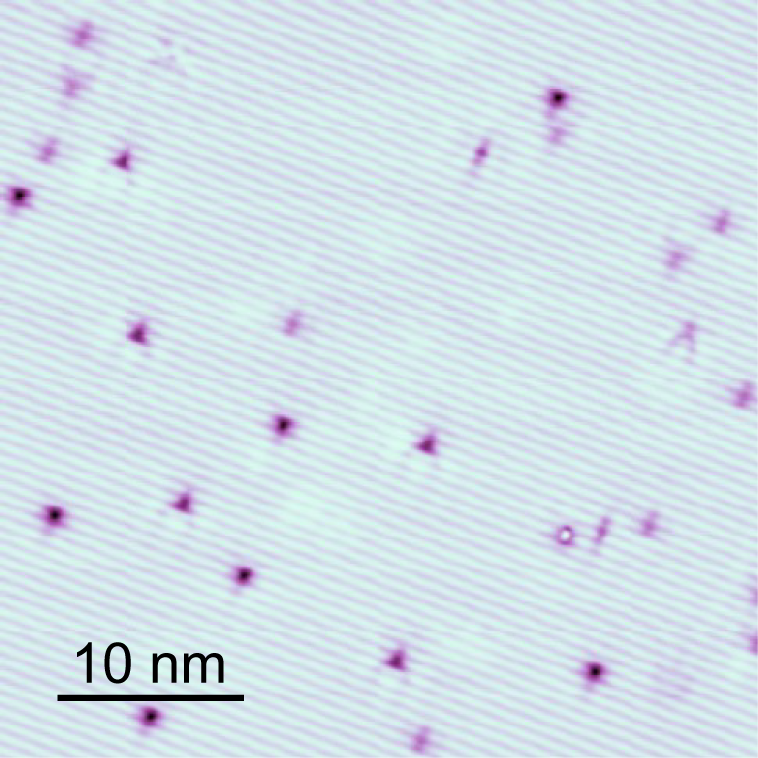}
\caption*{{\bf Fig.S4: Topography image for dI/dV mapping.} Setpoint: -80 mV, 50 pA. QPI is obtained by mapping dI/dV in this region. To generate enough defects for QPI measurement, the sample was firstly grown to be $\beta$-MoTe$_2$, then annealed in a furnace with thermal gradient from 1173 K to 873 K. At last, further annealing is performed in a 1223 K furnace.
}
\end{figure*}

\end{document}